\documentclass[11pt,reqno]{amsart}

\newtheorem{thm}{Theorem}[section]

\newtheorem{rmk}{Remark}[section]

\newcommand{\mysection}[1]{\section{#1}\setcounter{equation}{0}}

\newfont{\bb}{msbm10 at 12pt}

\def\pf{{\textit {Proof :} }}

\def\R{\hbox{\bb R}}

\def\N{\mathcal N}

\newcommand{\bal}{\begin{align}}      \newcommand{\eal}{\end{align}}
\newcommand{\ba}{\begin{array}}      \newcommand{\ea}{\end{array}}
\newcommand{\bc}{\begin{center}}     \newcommand{\ec}{\end{center}}
\newcommand{\be}{\begin{enumerate}}  \newcommand{\ee}{\end{enumerate}}
\newcommand{\beq}{\begin{eqnarray}}  \newcommand{\eeq}{\end{eqnarray}}
\newcommand{\beQ}{\begin{eqnarray*}} \newcommand{\eeQ}{\end{eqnarray*}}
\newcommand{\bi}{\begin{itemize}}    \newcommand{\ei}{\end{itemize}}
\newcommand{\bt}{\begin{tabular}}    \newcommand{\et}{\end{tabular}}
\newcommand{\bdm}{\begin{displaymath}} \newcommand{\edm}{\end{displaymath}}

\def\qed{\hfill{Q.E.D.}\smallskip}

\newcommand{\ls}{\setlength{\baselineskip}{12pt}
                 \setlength{\parskip}{3mm}}

\begin{document}

\title[ADM and Bondi energy-momenta III]{On the relation between ADM
and Bondi energy-momenta III -- perturbed radiative spatial
infinity}

\author{Wen-ling Huang}
\address[Wen-ling Huang]{Department Mathematik, Schwerpunkt GD, Universit\"at Hamburg, Bundesstr. 55,
D-20146 Hamburg, Germany} \email{huang@math.uni-hamburg.de}
\author{Xiao Zhang}
\address[Xiao Zhang]{Institute of Mathematics, Academy of Mathematics and
Systems Science, Chinese Academy of Sciences, Beijing 100080,
China}
\email{xzhang@amss.ac.cn}

\begin{abstract}
In a vacuum spacetime equipped with the Bondi's radiating metric
which is asymptotically flat at spatial infinity including
gravitational radiation ({\bf Condition D}), we establish the
relation between the ADM total energy-momentum and the Bondi
energy-momentum for perturbed radiative spatial infinity. The
perturbation is given by defining the ``real" time the sum of the
retarded time, the Euclidean distance and certain function $f$.
\end{abstract}


\maketitle \pagenumbering{arabic}

\mysection{Introduction}
 \ls

It is a fundamental problem in gravitational radiation what the
relation is between the ADM energy-momentum and the Bondi
energy-momentum. Under certain asymptotic flatness conditions at
spatial infinity, it was shown the ADM energy-momentum at spatial
infinity is the past limit of the Bondi energy-momentum in certain
vacuum spacetimes \cite{AM, H, V1, V2, CK, Z4}. However, it is
presumably believed the assumed asymptotic flatness at spatial
infinity in all above works precludes gravitational radiation, at
least near spatial infinity.

In \cite{Z4}, the second author introduced a weaker assumption of
asymptotic flatness at spatial infinity in Bondi's radiating
spacetimes ({\bf Condition D}). This condition can be viewed as
Sommerfeld's radiation condition at spatial infinity \cite{Z4}.
And it should not preclude gravitational radiation. Under this
condition, it is found that the ADM total energy of a $t$-slice is
no longer the past limit of the Bondi mass, and they differ by a
quantity related to the {\it news functions}. In \cite{HZ}, the
authors established the relation between the ADM total linear
momentum and the past limit of the Bondi ``momentum'' under {\bf
Condition D}. It is surprising that in this case the second
fundamental form $h$ on the $t$-slice falls off as
$O(\frac{1}{r})$, however, some nice cancellation occurs and the
ADM total linear momentum is still finite. The ADM total linear
momentum at spatial infinity is no longer the past limit of the
Bondi ``momentum'', and the difference is calculated.

The definition of the ``real'' time $t$ is essential in Bondi's
radiating spacetimes and the spatial infinity is defined as a $t$
slice. In the works of \cite{Z4, HZ}, the ``real'' time $t$ is
defined as the sum of the retarded coordinate $u$ and the
Euclidean distance $r$. However, this definition is rather
restricted, it does not hold true even in the Schwarzschild
spacetime. In this paper, we will perturb the ``real'' time $t$ by
adding certain function $f$. The spatial infinity is also defined
in terms of this new $t$-slice. We establish the relation between
the ADM energy-momentum and the Bondi energy-momentum under {\bf
Condition D}.

We do not consider the polyhomogeneous Bondi expansions in the
present paper, as studied in \cite{CMS}. In \cite{CMS} (Appendix
D), the logarithmic singularities at null infinity can be removed
in the axisymmetric case if the free function $\gamma _2 (u, x
^\alpha)$ is chosen to be zero and $\gamma _{3,1} (u _0, x
^\alpha)$ is chosen to be zero for some $u _0$. We believe this is
a generic property that under certain class of Bondi's radiating
metrics which make the ADM total energy-momentum and the Bondi
energy-momentum well-defined, the standard Bondi expansions
without any logarithmic singularity at null infinity can always be
achieved.

The paper is organized as follows: In Section 2, we state some
well-known results of Bondi, van der Burg, Metzner and Sachs. In
Section 3, we define the ``real'' time $t$ via adding a perturbation
$f$. We study the asymptotic behaviors of the induced metric and the
the second fundamental form of a $t$ slice. In Section 4, we derive
the difference between the ADM total energy and the past limit of
the Bondi mass. Under certain conditions, we can find a good
perturbation such that the ADM total energy equals the past limit of
the Bondi mass. In Section 5, we derive the difference between the
ADM total linear momentum and the past limit of the Bondi
``momentum''. We also find certain condition ensuring the existence
of good perturbation such that the ADM total linear momentum equals
the past limit of the Bondi ``momentum''. However, we find, in
general, it is impossible to perturb spatial infinity so that both
the ADM total energy and the ADM total linear momentum are the past
limit of the Bondi energy-momentum.

\mysection{Bondi's radiating vacuum spacetimes} \ls

The Bondi's radiating vacuum spacetime $\big(L ^{3,1}, \tilde
g\big) $ is a vacuum spacetime equipped with the following metric
$\tilde g =\tilde g _{ij} dx ^i dx ^j$
 \beq
\tilde g &=&\Big(\frac{V}{r} e ^{2\beta} +r ^2 e ^{2 \gamma} U ^2
\cosh 2\delta +r ^2 e ^{-2 \gamma} W ^2 \cosh 2\delta \nonumber\\
&&+2 r ^2 UW \sinh 2 \delta \Big)du ^2 -2e ^{2\beta}
du dr   \nonumber\\
& &-2r ^2 \Big(e ^{2 \gamma} U \cosh 2\delta +W \sinh 2 \delta
\Big) du d\theta     \nonumber\\
& &-2r ^2 \Big(e ^{-2 \gamma} W \cosh 2\delta+U \sinh 2\delta
\Big)\sin \theta du d\psi    \nonumber\\
& &+r ^2 \Big(e ^{2 \gamma} \cosh 2\delta d\theta ^2 +e ^{-2
\gamma}\cosh 2\delta \sin ^2 \theta d \psi ^2   \nonumber\\
& &+2 \sinh 2\delta \sin \theta d \theta d \psi \Big),
\label{bondi-metric}
 \eeq
where $\beta, \gamma, \delta, U, V, W$ are functions of
 \beQ
x ^0 =u,\;\;x ^1=r,\;\;x ^2=\theta, \;\;x ^3=\psi,
 \eeQ
$u$ is a retarded coordinate, $r$ is Euclidean distance, $\theta $
and $\psi $ are spherical coordinates, $0 \leq \theta \leq \pi$,
$0 \leq \psi \leq 2 \pi$. We assume that $\tilde g$ satisfies {\it
the outgoing radiation condition}.

The metric (\ref{bondi-metric}) was studied by Bondi, van der
Burg, Metzner and Sachs in the theory of gravitational waves in
general relativity \cite{BBM, S, vdB}. They proved that the
following asymptotic behavior holds for $r$ sufficiently large if
the spacetime satisfies the outgoing radiation condition
\cite{vdB}
 \beQ
\gamma &=&\frac{c(u, \theta, \psi)}{r} +\frac{C(u,\theta,
\psi)-\frac{1}{6} c ^3 -\frac{3}{2} c d ^2}{r ^3}
+O\Big(\frac{1}{r ^4}\Big),  \\
\delta &=&\frac{d(u, \theta, \psi)}{r} +\frac{H(u,\theta,
\psi)+\frac{1}{2} c ^2 d -\frac{1}{6} d ^3}{r ^3}
+O\Big(\frac{1}{r ^4}\Big),  \\
\beta &=&-\frac{c ^2 + d ^2}{4r ^2} +O\Big(\frac{1}{r ^4}\Big),     \\
U &=& -\frac{l(u, \theta, \psi)}{r ^2} +\frac{p(u, \theta,
\psi)}{r ^3}
+O\Big(\frac{1}{r ^4}\Big),  \\
W &=& -\frac{\bar l(u, \theta, \psi)}{r ^2}+\frac{\bar p(u,
\theta, \psi)}{r ^3}+O\Big(\frac{1}{r ^4}\Big),  \\
V &=& -r +2 M (u, \theta, \psi)+\frac{\bar M(u, \theta, \psi)}{r}
+O\Big(\frac{1}{r^2}\Big),
 \eeQ
where
 \beQ
l &=& c _{, 2} +2c \cot \theta +d _{, 3} \csc \theta,\\
\bar l &=& d _{, 2} +2d \cot \theta -c _{,3} \csc \theta,\\
p &=& 2N +3(c c _{, 2}+d d _{, 2}) +4(c ^2 +d ^2)\cot \theta\\
  & &-2(c_{,3} d -c d _{,3}) \csc \theta,\\
\bar p &=& 2P +2(c _{, 2} d -c d _{, 2}) +3(c c_{,3} +d d
_{,3})\csc \theta,\\
\bar M &=&N _{,2} +\cot \theta +P _{, 3} \csc \theta
-\frac{c ^2 +d ^2}{2}\\
& &-\big[(c _{,2}) ^2 +(d _{,2}) ^2 \big]-4(c c _{,2} +d d
_{,2})\cot \theta \\
& & -4(c ^2 +d ^2) \cot ^2
\theta -\big[(c _{,3}) ^2 +(d _{,3}) ^2 \big]\csc ^2 \theta \\
& &+4(c _{,3} d -c d _{, 3}) \csc \theta \cot \theta +2(c_{,3} d
_{,2}-c _{,2} d _{,3})\csc \theta.
 \eeQ
(We denote $f _{,\nu} = \frac{\partial f}{\partial x ^{\nu} }$ for
$\nu =0,1,2,3$ throughout the paper.) $M$ is the {\it mass aspect}
and $c_{,0}$, $d _{, 0}$ are the {\it news functions} and they
satisfy the following equation \cite{vdB}:
 \beq
M _{,0} =-\Big[(c _{,0} )^2 +(d _{, 0} )^2 \Big]
+\frac{1}{2}\Big(l _{,2} +l \cot \theta +\bar l _{,3} \csc \theta
\Big) _{,0} .\label{u-deriv}
 \eeq

Let $N _{u _0}$ be a null hypersurface which is given by $u=u _0$
at null infinity. The Bondi energy-momentum of $N _{u _0}$ is
defined by \cite{BBM}:
 \beQ
m _\nu (u _0) = \frac{1}{4 \pi} \int _{S ^2} M (u _0, \theta,
\psi) n ^{\nu} d S
 \eeQ
where $\nu =0, 1, 2, 3$, $S ^2 $ is the unit sphere,
 \beQ
n ^0 =1,\;\; n ^1 = \sin \theta \cos \psi,\;\; n ^2 = \sin \theta
\sin \psi,\;\; n ^3 = \cos \theta.
 \eeQ
And $m _0$ is the Bondi mass, $m _i$ is the Bondi momentum.

Denote by $\{ \breve{e} ^i \}$ the coframe of the standard flat
metric $g _0$ on $\R ^3$ in polar coordinates,
 \beQ
\breve{e} ^1 = dr, \;\;\breve{e} ^2 = r d\theta, \;\;\breve{e} ^3
= r \sin \theta d\psi.
 \eeQ
Denote by $\{\breve{e} _i \}$ the dual frame ($i=1,2,3$). The
connection 1-form $\{\breve{\omega } _{ij}\}$ is defined by
 \beQ
d \breve{e} ^i= - \breve{\omega} _{ij} \wedge \breve{e} ^{j}.
 \eeQ
It is easy to find that
 \beQ
    \breve{\omega}  _{12} =  -\frac{1}{r}\breve{e}  ^{2}, \;\;
    \breve{\omega}  _{13} =  -\frac{1}{r}\breve{e}  ^{3}, \;\;
    \breve{\omega}  _{23} =  -\frac{\cot \theta}{r}\breve{e} ^{3}.
 \eeQ
The Levi-Civita connection $\breve{\nabla}$ of $g _0$ is given by
 \beQ
\breve{\nabla} \breve{e} _{i}= - \breve{\omega} _{ij} \otimes
\breve{e} _{j}.
 \eeQ
We denote $\breve{\nabla} _i \equiv \breve{\nabla} _{\breve{e}
_i}$ for $i=1,2,3$ throughout the paper.

Define $\mathcal{C} _{\{a_1, a_2, a_3\}}$ the space of smooth
functions in the spacetime which satisfies the following
asymptotic behavior at spatial infinity
 \beq
\mathcal{C} _{\{a_1, a_2, a_3\}} =
 \left\{f:
 \begin{array}{ccc}
    \lim _{r \rightarrow \infty}\lim _{u \rightarrow -\infty} r ^{a_1}f
    &=&O\big(1\big),\\
    \lim _{r \rightarrow \infty}\lim _{u \rightarrow -\infty} r ^{a_2}
    \breve{\nabla} _i f
    &=&O\big(1\big),\\
    \lim _{r \rightarrow \infty}\lim _{u \rightarrow -\infty} r ^{a_3}
    \breve{\nabla} _i\breve{\nabla} _j f &=& O\big(1\big)
 \end{array}
 \right\}.       \label{mathcal-C}
 \eeq
In \cite{Z4}, the following four conditions were introduced:
 \begin{description}
 \item[Condition A] {\bf Each of the six functions $\beta $,
$\gamma$, $\delta$, $U$, $V$, $W$ together with its derivatives up
to the second orders are equal at $\psi =0$ and $2\pi$}.\\
 \item[Condition B] {\bf For all $u$,
 \beQ
\int _0 ^{2\pi} c(u, 0, \psi) d\psi =0, \;\; \int _0 ^{2\pi} c(u,
\pi, \psi) d\psi =0.
 \eeQ}
 \item[Condition C] $\;\;\gamma \in \mathcal{C} _{\{1, 2, 3\}},\;\;
\delta \in \mathcal{C} _{\{1, 2, 3\}}, \;\;\beta \in \mathcal{C}
_{\{2, 3, 4\}}, \;\;U \in \mathcal{C} _{\{2, 3, 4\}}, \;\;W \in
\mathcal{C} _{\{2, 3, 4\}}, \;\;V+r \in \mathcal{C} _{\{0, 1,
2\}}$.\\
 \item[Condition D] $\;\;\gamma \in \mathcal{C} _{\{1, 1,
1\}},\;\; \delta \in \mathcal{C} _{\{1, 1, 1\}}, \;\;\beta \in
\mathcal{C} _{\{2, 2, 2\}}, \;\;U \in \mathcal{C} _{\{2, 2, 2\}},
\;\;W \in \mathcal{C} _{\{2, 2, 2\}}, \;\;V+r \in \mathcal{C}
_{\{0, 0, 0\}}$.\\
 \end{description}

{\bf Condition A} and {\bf Condition B} ensure that the metric
(\ref{bondi-metric}) is regular, also ensure the following Bondi
mass loss formula
 \beQ
\frac{d}{du} m _\nu =-\frac{1}{4 \pi} \int _{S ^2} \big[(c _{,0}) ^2
+(d _{,0}) ^2 \big] n ^\nu d S.
 \eeQ
{\bf Condition C} ensures the Schoen-Yau's positive mass theorem
at spatial infinity. However, it precludes gravitational
radiation. {\bf Condition D} should include gravitational
radiation. It indicates that, for $r$ sufficiently large,
 \beQ
\lim _{u \rightarrow -\infty} \Big\{M, c, d, M _{,0}, c _{,0}, d
_{,0}, M _{,A}, c _{,A}, d _{,A} \Big\}=O\big(1\big)
 \eeQ
where $2 \leq A\leq 3$. We refer to \cite{Z4} for some physical
interpretation of {\bf Condition D}.
\mysection{Initial data sets} \ls

From now on, we assume the ``real'' time $t$ is defined as
 \beq
t=u+r+f(r, \theta, \psi) \label{real-t}
 \eeq
for $r$ sufficiently large, where $f$ is smooth function which has
the following asymptotic behavior
 \beq
f =a _1 \ln r +a _2(\theta, \psi) +a _{3}(r,\theta, \psi)
\label{f}
 \eeq
for $r$ sufficiently large, where $a_1$ is constant, $a _2$, $a_3$
are smooth functions which satisfy
 \beQ
a _i \big | _{\psi =0}=a _i \big | _{\psi =2\pi}, \;\;a _{i,A}
\big | _{\psi =0}=a _{i, A}\big | _{\psi =2\pi},\;\; a _{i,AB}
\big | _{\psi =0}=a _{i, AB}\big | _{\psi =2\pi}
 \eeQ
for $i=2,3$, $A, B=2,3$. Moreover,
 \beQ
 a_3 =O\Big(\frac{1}{r}\Big),\;\;
 \breve{\nabla} _k a_3
 =O\Big(\frac{1}{r^2}\Big),\;\;
 \breve{\nabla} _l \breve{\nabla} _k  a_3
 =O\Big(\frac{1}{r^3}\Big).
 \eeQ
for $r$ sufficiently large.

Substituting (\ref{real-t}) into (\ref{bondi-metric}), we obtain the
the spacetime metric
 \beQ
\tilde g =\tilde g _{tt} dt ^2 +2 \tilde g _{ti}dtdx ^i +\tilde g
_{ij}dx ^i dx ^j
 \eeQ
$(1 \leq i,j \leq 3)$. An initial data set $\big(N _{t _0}, g,
h\big)$ is a spacelike hypersurface in $L ^{3,1}$ which is given by
$\big\{t= t _0\big\}$. Here $g$ is the induced metric of $\tilde g$
and $h$ is the second fundamental form. The lapse $\N $ and the
shift $X _i$ $(i=1,2,3)$ of the spacelike hypersurface $N _{t _0}$
are
 \beQ
 \N &=&\Big(-\tilde g ^{tt} \Big) ^{-\frac{1}{2}}
 =\Big(-\tilde g _{tt}
+\tilde g _{ti} \tilde g _{tj} g ^{ij}\Big) ^{\frac{1}{2}},\\
X _i &=&\tilde g _{ti}.
 \eeQ
The second fundamental form is then given by
 \beQ
h _{ij} =\frac{1}{2 \N} \Big(\nabla _i X _j +\nabla _j X _i
-\partial _t \tilde g _{ij} \Big) _{t=t_0}.
 \eeQ

With the help of asymptotic behavior of $\beta, \gamma, \delta, U,
V, W, f$ and Mathematica 5.0, we obtain the asymptotic expansion
of $g _{ij}$ and $h _{ij}$:
 \beQ
g _{11}& =&1 + \frac{2\,M}{r}+\frac{1}{r^2}\Big[
-\frac{{c}^2}{2} - \frac{{d}^2}{2} +l ^2 +{\bar l }^2 +\bar M -a _1 ^2 -4M a _1 \Big]\\
       & &+O\Big(\frac{1}{r ^3}\Big),\\
g _{22}& =&r^2 + 2 r c + 2 {c}^2 + 2 {d}^2
       +2 l a _{2,2} -a _{2,2} ^2
           +O\Big(\frac{1}{r}\Big),\\
 g _{33}& =&\Big(r^2  - 2 r c  + 2c ^2 +
  2 d ^2 \Big)\sin ^2 \theta +2 \bar l \sin \theta a _{2,3}-a _{2,3} ^2 +O\Big(\frac{1}{r}\Big),\\
  g_{12}&=&-l + \frac{1}{r}\Big( -2c l -2d \bar l
  +p +a_{1} l -a _1 a _{2,2} -2M a _{2,2}\Big)
+O\Big(\frac{1}{r ^2}\Big) ,\\
   g _{13}&=&-\bar l \sin \theta\
   + \frac{1}{r}\Big[ \big(2 c \bar l   -2d l+\bar p \big)\sin \theta +\bar l
\sin \theta a_{1} -a _1 a _{2,3} -2M a
   _{2,3} \Big]\\
   & &+O\Big(\frac{1}{r ^2}\Big),\\
   g _{23}&=& 2 r d \sin \theta +l a _{2,3}+\bar l \sin \theta a _{2,2} -a _{2,2} a_{2,3}
+O\Big(\frac{1}{r}\Big),\\
 h _{11}& =&\frac{M _{,0}}{r}
+\frac{1}{r ^2}\Big(2M-M M _{,0} + \frac{c c _{,0}}{2}+\frac{d d
_{,0}}{2}+l l _{,0}+\frac{\bar M
_{,0}}{2}\\
& &+a _1 -3 M _{,0} a _1+\bar l _{,0}\csc \theta a _{2,3} \Big)+O\Big(\frac{1}{r ^3}\Big),\\
h _{22}& =&-r c _{,0}-2 M+M c _{,0} -2c c _{,0} -2 d d_{,0}+l
 _{,2}-a _1 -a _{2,22} +c _{,0} a _1\\
 &&+O\Big(\frac{1}{r}\Big),\\
 h _{33}& =& \Big(r c_{,0} -2 M  - M c _{,0} -2 c c _{,0}
 -2 d d_{,0} +l \cot \theta + \bar l
 _{,3} \csc \theta\Big)\sin ^2 \theta \\
 & &- \sin ^2 \theta a _1 -a _{2,33}-\sin \theta \cos
 \theta a _{2,2}
-  c _{,0} \sin ^2 \theta a _1 +O\Big(\frac{1}{r}\Big),\\
 h _{12}&=&\frac{1}{r}\Big[- M_{,2} -l + c_{,0} l+d _{,0} \bar l
-\big(c _{,0}+M _{,0}-1\big)a _{2,2} -d _{,0} \csc \theta a _{2,3}\Big]\\
& &+O\Big(\frac{1}{r ^2}\Big),\\
 h _{13}&=&\frac{1}{r}\Big[\big(-M_{,3}\csc \theta -\bar l  - c_{,0} \bar l   + d
 _{,0}l\big)\sin \theta + \big(c _{,0}-M _{,0}+1\big)a _{2,3}\\
 & & -d _{,0} \sin \theta a _{2,2}\Big]+O\Big(\frac{1}{r ^2}\Big) ,\\
 h _{23}&=&\Big[-
  r d_{,0}+M d_{,0}+\frac{1}{2}\big( \bar l _{,2}-\bar l
  \cot \theta +l _{,3} \csc \theta \big)\Big]\sin \theta -a _{2,23}\\
  & &+d _{,0} \sin \theta a _1+\cot \theta a
 _{2,3}+O\Big(\frac{1}{r}\Big).
 \eeQ
The trace of the second fundamental form is
 \beQ
tr _g \big(h\big)&=&\frac{M_{,0}}{r} + \frac{1}{r^2}\Big[-2M -3M M
_{,0} +l _{,2} +l \cot \theta +\bar
l _{,3} \csc\theta \\
& &+\frac{c c_{,0}}{2}+\frac{d d_{,0}}{2} +l l _{,0} +\frac{\bar M
_{,0}}{2} -\big(a _1 + 3 M_{,0}a _1+a _{2, 22}\\
& &+ \csc ^2 \theta a _{2, 33} +\cot \theta a _{2,2}\big) +\bar l
_{,0}\csc \theta a _{2,3} \Big]
     +O\Big(\frac{1}{r ^3}\Big).
 \eeQ
\mysection{ADM and Bondi total energy} \ls

In \cite{Z4, HZ}, the authors derived the relation between the ADM
total energy, the total linear momentum and the Bondi
energy-momentum under {\bf Condition A}, {\bf Condition B} and
{\bf Condition D} and under the assumption $t=u+r$. In this
section, we study the relation between them under these conditions
and under (\ref{real-t}) with condition (\ref{f}). Let Euclidean
coordinates
 \beQ
y ^1 =r \sin \theta \cos \psi,\;\;\;\; y ^2=r \sin \theta \sin
\psi, \;\;\;\;y ^3 =r \cos \theta.
 \eeQ
In polar coordinates, the ADM total energy $\mathbb{E}$ and the
ADM total linear momentum $\mathbb{P} _k$ are \cite{ADM, Z4, HZ}
 \beQ
 \mathbb{E} &=&\frac{1}{16\pi}\lim _{r \rightarrow \infty} \int
_{S _r} \Big[\breve {\nabla } ^j g\big(\breve{e} _1, \breve{e}
_j\big) -\breve {\nabla } _1 tr _{g _0} \big(g\big) \Big]
\breve{e} ^2 \wedge \breve{e} ^3,\\
 \mathbb{P} _k &=&\frac{1}{8\pi}\lim _{r \rightarrow
\infty} \int _{S _r} \Big[h\Big(\frac{\partial}{\partial y ^k},
\frac{\partial}{\partial r}\Big) -g\Big(\frac{\partial}{\partial y
^k}, \frac{\partial}{\partial r}\Big)tr _{g} \big(h\big) \Big]
\breve{e} ^2 \wedge \breve{e} ^3.
 \eeQ
 \begin{thm}\label{energy}
Let $\mathbb{E}(t _0)$ be the ADM total energy of the initial data
set $\big(N _{t_0}, g, h \big)$ where $t$ is given by
(\ref{real-t}) with condition (\ref{f}). Under {\bf Condition A},
{\bf Condition B} and {\bf Condition D}, we have
 \beQ
\mathbb{E} (t _0)& = &m _0 (-\infty) +\frac{1}{2\pi}\lim
_{u\rightarrow -\infty} \int _0 ^{\pi} \int _0 ^{2\pi} \big(c c_{,0}
+d d _{,0} \big) \sin\theta d\psi d\theta\\ & &+\frac{1}{16\pi}\lim
_{u\rightarrow -\infty} \int _0 ^{\pi} \int _0 ^{2\pi}\big(a
_{2,2}\, \sin \theta \,l _{,0} +a _{2,3} \bar l _{,0}\big) d\psi
d\theta.
 \eeQ
 \end{thm}
\pf Under {\bf Condition A}, {\bf Condition B} and {\bf Condition
D}, we have, as $r \rightarrow \infty$, (or $u \rightarrow
-\infty$),
 \beQ
\breve {\nabla } ^j g\big(\breve{e} _1, \breve{e} _j\big) -\breve
{\nabla } _1 tr _{g _0} \big(g\big)
 &=&\breve {e} _j \Big(g\big(\breve{e} _1, \breve{e} _j\big)\Big)
 -\breve{e} _1 tr _{g _0} \big(g\big)\\
 & &-g\big(\breve{e} _j, \breve{e} _i\big)\breve {\omega} _{i1}
 \big(\breve{e} _j \big)-g\big(\breve{e} _1, \breve{e} _i\big)\breve {\omega} _{ij}
 \big(\breve{e} _j \big)\\
 &=&\frac{4M}{r ^2}-\frac{l _{,2}}{r ^2} -\frac{l \cot \theta }{r
 ^2}-\frac{\bar l _{,3}}{r ^2 \sin \theta}
 -4\breve{e} _1\Big(\frac{c ^2 +d ^2}{r ^2} \Big)\\
 & &+ \frac{a _{2,2}\,
l _{,0} +a _{2,3} \csc \theta \bar l _{,0}}{r ^2}
+O\Big(\frac{1}{r ^3}\Big)\\
 &=&\frac{4M}{r ^2} -\frac{l _{,2}}{r ^2} -\frac{l \cot \theta }{r
 ^2}-\frac{\bar l _{,3}}{r ^2 \sin \theta}+\frac{4\big(c ^2 +d ^2\big) _{,0}}{r ^2} \\
 & &+ \frac{a _{2,2}\, l
_{,0} +a _{2,3} \csc \theta \bar l _{,0}}{r ^2}+O\Big(\frac{1}{r
^3}\Big).
 \eeQ
Therefore the theorem is a direct consequence of integrating it
over $S _r$ and using that for fixed $t=t _0$, $r \rightarrow
\infty$ is equivalent to $u \rightarrow -\infty$. \qed
\begin{rmk}
If $a_2$ is chosen to be a constant, the difference $\mathbb{E}
(t_0)- m _0 (-\infty)$ is independent on the choice of $f$, which
is invariant in the perturbed class (\ref{f}) that $a_1$, $a_2$
are constant.
\end{rmk}

The following theorem indicates that, under certain conditions, we
can choose suitable $f$ such that, at spatial infinity defined by
(\ref{f}), the ADM total energy is the past limit of the Bondi
mass.

 \begin{thm}\label{energy-f}
Suppose that {\bf Condition A}, {\bf Condition B} and {\bf
Condition D} hold in the Bondi's radiating spacetime
(\ref{bondi-metric}). If
 \beq
\lim _{u \rightarrow -\infty} c _{,0} \big| _{\theta =0} &= &\lim
_{u
\rightarrow -\infty} c _{,0} \big| _{\theta =\pi } =0,\label{cond1}\\
\lim _{u \rightarrow -\infty} d _{,30} \big| _{\theta =0} &= &\lim
_{u
\rightarrow -\infty} d _{,30} \big| _{\theta =\pi } =0,\label{cond2}\\
\lim _{u \rightarrow -\infty} \big[M_{,0} +(c _{,0} ) ^2 +(d _{,0} )
^2 \big]&\neq &0 \label{cond3},
 \eeq
then there exists $f_0$ such that
 \beQ
\mathbb{E} _{f_0} (t _0) = m _0 (-\infty)
 \eeQ
where $\mathbb{E} _{f _0} (t _0)$ is the ADM total energy of the
initial data set $\big(N _{t_0}, g, h \big)$ where $t$ is given by
(\ref{real-t}) and $f=f_0$ given by (\ref{f}) with
 \beq
a _2  = \lim _{u \rightarrow -\infty} \frac{4c c _{,0} +4d d
_{,0}}{M _{,0} +(c _{,0}) ^2 +(d _{,0}) ^2 }. \label{a2}
 \eeq
 \end{thm}
\pf Note that
 \beQ
\lim _{u \rightarrow -\infty} \int _0 ^{\pi}  a _{2,2} \sin \theta
l _{,0} d\theta &=&\lim _{u \rightarrow -\infty} a _{2}\big(c
_{,20} \sin \theta +2 c _{,0} \cos \theta +d _{,30}\big) \Big| _0
^{\pi}\\
 & &-\lim _{u \rightarrow -\infty} \int _0 ^{\pi}
 a _{2} \big(\sin \theta l _{,0}\big) _{,2} d\psi d\theta \\
 &=&-\lim _{u \rightarrow -\infty} \int _0 ^{\pi}
 a _{2} \big(\cos \theta l _{,0}+\sin \theta l _{,02}\big) d\theta , \\
\lim _{u \rightarrow -\infty} \int _0 ^{2\pi} a _{2,3} {\bar l}
_{,0} d\psi
 &=& \lim _{u \rightarrow -\infty} a _{2}\big(d _{,20}+2 d _{,0} \cot \theta -c
_{,30} \csc \theta \big) \Big| ^{2\pi} _0  \\
 & &- \lim _{u \rightarrow -\infty} \int _0 ^{2\pi} a _{2} \bar l _{,03} d\psi\\
 &=&- \lim _{u \rightarrow -\infty} \int _0 ^{2\pi} a _{2} \bar l _{,03} d\psi.
 \eeQ
We obtain
 \beQ
 & &\lim _{u \rightarrow -\infty} \int _0 ^{\pi} \int _0
^{2\pi}\big(a _{2,2}\, \sin \theta \,l _{,0} +a _{2,3} \bar l
_{,0}\big) d\psi d\theta\\
 &=& -\lim _{u \rightarrow -\infty}
\int _0 ^{\pi} \int _0 ^{2\pi}a _2 \big(l _{,2} +l \cot \theta
+\bar l _{,3} \csc \theta \big) _{,0} \sin \theta d\psi d\theta \\
 &=&-2 \lim _{u \rightarrow -\infty} \int _0 ^{\pi} \int _0 ^{2\pi}
 a _2 \big[M _{,0} +(c _{,0}) ^2 +(d _{,0}) ^2 \big] \sin \theta d\psi d\theta .
 \eeQ
Therefore the theorem is a direct consequence of Theorem
\ref{energy}. \qed
\mysection{ADM and Bondi total momenta} \ls

In this section, we study the relation between the ADM total
linear momentum and the Bondi momentum.

 \begin{thm}\label{momentum}
Let $\mathbb{P} _k (t _0)$ be the ADM total linear momentum of the
initial data set $\big(N _{t_0}, g, h \big)$ where $t$ is given by
(\ref{real-t}) with condition (\ref{f}). Under {\bf Condition A},
{\bf Condition B} and {\bf Condition D}, we have
 \beQ
\mathbb{P} _k (t_0)= m _k (-\infty)+\frac{1}{8\pi}\lim _{u
\rightarrow -\infty} \int _0 ^{\pi} \int _0 ^{2
\pi}\overline{{\mathcal P}} _k d \psi d \theta
 \eeQ
for $k=1,2,3$, where where
 \beQ
\overline{{\mathcal P}} _1 &=&\Big[\big(c _{,0}+M _{,0}\big) \cos
\theta
\cos \psi -d_{,0} \sin \psi\Big]\big(l -a _{2,2}\big)\sin \theta\\
             & &+\Big[\big(c _{,0}-M _{,0}\big)
\sin \psi +d_{,0} \cos \theta \cos \psi\Big]\big(\bar l \sin \theta -a _{2,3}\big),\\
\overline{{\mathcal P}} _2 &=&\Big[\big(c _{,0}+M _{,0}\big) \cos
\theta \sin \psi +d_{,0} \cos \psi\Big]\big(l -a _{2,2}\big) \sin \theta \\
             & &-\Big[\big(c _{,0}-M _{,0}\big)
\cos \psi -d_{,0} \cos \theta \sin \psi\Big]\big(\bar l \sin \theta -a _{2,3}\big),\\
\overline{{\mathcal P}} _3 &=&-\big(c _{,0}+M _{,0}\big)\big(l -a
_{2,2}\big) \sin ^2 \theta  - d _{,0} \big(\bar l \sin \theta -a
_{2,3}\big)\sin \theta.
 \eeQ
 \end{thm}
\pf Denote $\mathbb{K} _k =h\big(\frac{\partial}{\partial y ^k},
\frac{\partial}{\partial r}\big) -g\big(\frac{\partial}{\partial y
^k}, \frac{\partial}{\partial r}\big)tr _{g} \big(h\big)$. Using the
asymptotic expansions of $g _{ij}$ and $h _{ij}$, we obtain
 \beQ
\mathbb{K} _1&=&\frac{1}{r ^2}\Big\{4M \sin \theta \cos \psi -M
_{,2}
\cos \theta \cos \psi +M _{,3} \csc \theta \sin \psi \\
& &- l _{,2} \sin \theta \cos \psi-\bar l _{,3} \cos \psi \\
& &+ \big[( c _{,0}+M _{,0}-2) \cos \theta \cos \psi -d _{,0} \sin
\psi  \big] (l-a_{2,2}) \\
& & + \big[(c _{,0} -M _{,0}+1) \sin \psi +d _{,0} \cos \theta \cos
\psi \big](\bar l -\csc\theta a _{2,3})\\
& &+2 \sin \theta \cos \psi a _1
  +\csc \theta \cos \psi  a _{2,33} +
  \sin \theta \cos \psi  a _{2,22}\Big\}
  +O\Big(\frac{1}{r ^3}\Big),\\
\mathbb{K} _2&=&\frac{1}{r ^2}\Big\{4M \sin \theta \sin \psi -M
_{,2}\cos \theta \sin \psi -M _{,3} \csc \theta \cos \psi\\
& &- l _{,2} \sin \theta \sin \psi -\bar l _{,3} \sin
\psi\\
& &+\big[( c _{,0} +M _{,0}-2)\cos \theta \sin \psi +d _{,0} \cos
\psi  \big](l-a_{2,2}) \\
& & - \big[(c _{,0} -M _{,0}+1) \cos \psi -d _{,0} \cos \theta \sin
\psi \big](\bar l -\csc \theta a_{2,3})\\
& &+2 a _1 \sin \theta \sin \psi +
  \csc \theta \sin \psi a _{2,33} +
  \sin \theta \sin \psi a _{2,22}\Big\}+O\Big(\frac{1}{r ^3}\Big).\\
\mathbb{K} _3&=&\frac{1}{r ^2}\Big\{4M \cos \theta  -M _{,2}
\sin \theta - l _{,2} \cos \theta -\bar l _{,3} \cot \theta  \\
& &-\big[( c _{,0} +M _{,0}-2)\sin \theta  +\csc \theta \big]l -d
_{,0} (\bar l \sin \theta -a_{2,3})\\
& &+2 a_1 \cos \theta +\cos \theta
\cot \theta a_{2,22}+\cot \theta \csc \theta a _{2,33}\\
& &+ \big(c _{,0}+M _{,0}-1\big)\sin \theta a
_{2,2}\Big\}+O\Big(\frac{1}{r ^3}\Big).
 \eeQ
Since $a _1$ is constant,
 \beQ
\int _{S ^2} a _1 n ^i dS =0
 \eeQ
for $i=1,2,3$, then the theorem is a direct consequence of
integrating $\mathbb{K} _k$ over $S _r$ and using that for fixed
$t=t _0$, $r \rightarrow \infty$ is equivalent to $u \rightarrow
-\infty$.\qed

\begin{rmk}
If $a_2$ is chosen to be a constant, then the difference $\mathbb{P}
_k (t_0)- m _k (-\infty)$ are independent on the choice of $f$,
which are invariant in the perturbed class (\ref{f}) that $a_1$,
$a_2$ are constant.
\end{rmk}

 \begin{thm}\label{momentum-f}
Suppose that {\bf Condition A}, {\bf Condition B} and {\bf Condition
D} hold in the Bondi's radiating spacetime (\ref{bondi-metric}). If
 \beq
\lim _{u \rightarrow -\infty}\big(\bar l _{,2} +\bar l \cot \theta
-l _{,3} \csc \theta \big)=0,  \label{cond4}
 \eeq
then there exists $f_1$ such that
 \beQ
\mathbb{P} _{k,f_1} (t _0) = m _k (-\infty)
 \eeQ
where $\mathbb{P} _{k, f _1} (t _0)$ is the ADM total linear
momentum of the initial data set $\big(N _{t_0}, g, h \big)$ where
$t$ is given by (\ref{real-t}) and $f=f_1$ given by (\ref{f}) with
$a _2$ satisfying
 \beq
a _{2,2}=l,\;\;\;\;a _{2,3} =\bar l \sin \theta . \label{aa2}
 \eeq
 \end{thm}
 \pf The condition (\ref{cond4}) ensures the existence of
 (\ref{aa2}). Therefore the theorem follows. \qed

\begin{rmk}
From Theorem \ref{energy-f} and Theorem \ref{momentum-f}, we know
that, in general, it is impossible to perturb spatial infinity via
the function $f$ so that both the ADM total energy and the ADM total
linear momentum are the past limit of the Bondi energy-momentum.
\end{rmk}

\hspace{2cm}

{\footnotesize {\it Acknowledgements.} {Xiao Zhang is partially
supported by National Natural Science Foundation of China under
grants 10231050, 10421001 and the Innovation Project of Chinese
Academy of Sciences. This work was partially done when Wen-ling
Huang visited the Morningside Center of Mathematics, Chinese Academy
of Sciences, and she would like to thank the center for its
hospitality.}}

\end{document}